\documentclass[prl,aps, english, twocolumn, hyperref,superscriptaddress]{revtex4}

\usepackage{graphicx}

\usepackage{amssymb, amsmath, color, rotating}
\newcommand{\J}{{t}}
\newcommand{\tb}{{\tt t}}

\begin{document}

\title{Evolution of condensate fraction during rapid lattice ramps}

\author{Stefan S. Natu}

\email{ssn8@cornell.edu}

\affiliation{Laboratory of Atomic and Solid State Physics, Cornell University, Ithaca, NY 14853, USA.}

\author{David C. McKay}

\affiliation{Department of Physics, University of Illinois, 1110 W Green St., Urbana, IL 61801}

\author{Brian DeMarco}

\affiliation{Department of Physics, University of Illinois, 1110 W Green St., Urbana, IL 61801}

\author{Erich J. Mueller}

\affiliation{Laboratory of Atomic and Solid State Physics, Cornell University, Ithaca, NY 14853, USA.}

\begin{abstract}
Combining experiments and numerical simulations, we investigate the redistribution of quasi-momentum in a gas of atoms trapped in an optical lattice when the lattice depth is rapidly reduced.  We find that interactions lead to significant momentum redistribution on millisecond timescales, thereby invalidating previous assumptions regarding adiabaticity. We show that this phenomenon is driven by the presence of low-momentum particle-hole excitations in an interacting system. Our results invalidate bandmapping as an equilibrium probe in interacting gases. \end{abstract}
\maketitle

\textit{Introduction---}
Optical lattice experiments are making important contributions to our knowledge of equilibrium and non-equilibrium properties of quantum many-body systems \cite{lewensteinreview}. 
A prime example is that of bosons in optical lattices \cite{bosehub}, which are described by the Bose Hubbard Hamiltonian. There are many of the open questions in this system related to transport and dynamics
\cite{noneqrefs}. 
Here we explore one such issue: how does the condensate fraction evolve during a lattice ramp?

In addition to its relevance for developing paradigms of nonequilibrium physics in cold gases, 
our study is of practical value: lattice ramps are routinely used in a probe known as {\em bandmapping}  \cite{greiner, phillips, kohl}.
The bandmapping protocol involves three steps.
 First, the lattice depth is reduced over a carefully chosen timescale $\tau$, with the intention of mapping quasi-momentum onto momentum.  Second, the gas is allowed to ballistically expand for a short time,  mapping momentum onto position \cite{muellertof}.  Third, the gas is imaged.  These images have been interpreted as the initial quasi-momentum distribution. 
 This protocol has been used to measure condensate fraction \cite{bdm}, map Brillouin zones \cite{bloch},
determine temperatures \cite{bdm2}, and probe phase transitions \cite{spielman, bloch2, bdm3}. 
Here we show that bandmapping is an 
 inherently non-equilibrium process.  Understanding the timescale of evolution of quasi-momentum is therefore crucial to interpreting the data.  
 We find that in interacting systems the dynamics are much faster than previously believed, leading to systematic errors in measurements of quantities such as condensate fraction.
 
The key step in bandmapping is turning off the lattice slowly enough such that the quasi-momentum states adiabatically evolve into momentum states, yet fast enough to leave the occupations of different quasi-momentum states unchanged. For 
harmonically trapped
non-interacting particles, there are three relevant energy scales: the bandgap $E_{bg}$, the tunneling $\J$, and the quantum energy of the harmonic confining potential $h\nu$ ($h=2\pi\hbar$ is Planck's constant).  Provided that the lattice ramp time $\tau$ is long compared to the inverse bandgap ($\tau\gg h/E_{bg}$), the quasi-momentum states will adiabatically evolve into momentum states \cite{bdm2}.  To avoid motion of atoms in the trap, the lattice must be turned off quickly compared to the trap period ($\tau\ll 1/\nu$).  Thus, $h/E_{bg}$ and $1/\nu$ set the adiabatic and diabatic timescales respectively.  Aside from its influence on $E_{bg}$ and $\nu$ \cite{rey}, the hopping rate $\J$ has little direct impact on bandmapping.  In standard experiments in the tight-binding regime, the natural separation of timescales between $h/E_{bg}\lesssim0.1$~ms and $1/\nu\gtrsim 10$~ms makes it straightforward to satisfy the adiabatic condition; a ramp time $\tau\sim 1$~ms is usually employed.

Complications arise for an interacting gas. Interactions lead to a coherent redistribution of  quasi-momentum occupations \cite{natuinprep, rigol}.  Furthermore, collisions scatter atoms between different quasi-momenta while conserving total quasi-momentum. The on-site interaction energy between two atoms $U$ determines the relevant dynamical timescale $h/U$; in most experiments, $h/U\lesssim1$~ms. We show that the additional criterion, $\tau\ll h/U$, disrupts the separation of timescales that makes bandmapping successful in noninteracting systems.

Using a combination of experiment and numerical simulations, we investigate the impact of interactions on bandmapping for atoms confined in a lattice in the strongly correlated regime.  We quantify the redistribution of  quasi-momentum during lattice ramps for a Bose-Einstein condensate of atoms in a 3D cubic optical lattice.  The fraction of atoms in the condensate is determined after linearly ramping from lattice depth $V_i$ (with $10E_R<V_i<14E_R$, spanning the superfluid and Mott-insulator regimes) to a fixed final depth $V_f=4E_R$.  Here $E_R=(h/\lambda)^2/2m$ is the recoil energy,  $\lambda$ is the laser wavelength, and  $m$ is the atomic mass. 
The final depth is chosen so that the atoms remain in the single-band Bose-Hubbard limit, thereby simplifying comparison to numerical simulations. Ramps terminating at $V_f=0$ produce similar results.  Our results apply to interacting gases in general, including fermionic systems and mixtures.

\textit{Experimental Method.---} Our experimental setup is described in detail in Ref. \cite{bdm2}. In summary, we create a condensate composed of $^{87}$Rb atoms in the $|F=1,m_F=-1\rangle$ state in a harmonic trap with (geometric) mean trap frequency $\bar{\nu}_0=35.78(6)$~Hz.  We cool the gas 
until the condensate fraction exceeds $80\%$.

  We superimpose a cubic optical lattice with a $d=\lambda/2=406$~nm lattice spacing  on the atoms by slowly turning on three pairs of retro-reflected laser beams.  The laser intensity determines the potential depth $V$.  Through Kapitza-Dirac diffraction, we calibrate $V$ to within 1\%, but drift in the calibration results in a 6\% systematic uncertainty.  The Gaussian envelope of the lattice beams adds to the harmonic confinement, and the overall (geometric) mean trap frequency with the lattice on is $\bar{\nu} \approx \sqrt{\bar{\nu}_0^2 + \frac{8V_i}{(2 \pi)^2 m w^2}}$, where $w=(120\pm10)$~$\mu$m is the measured $1/e^2$ radius of the lattice laser beams.

\begin{figure}
\includegraphics[scale=0.6]{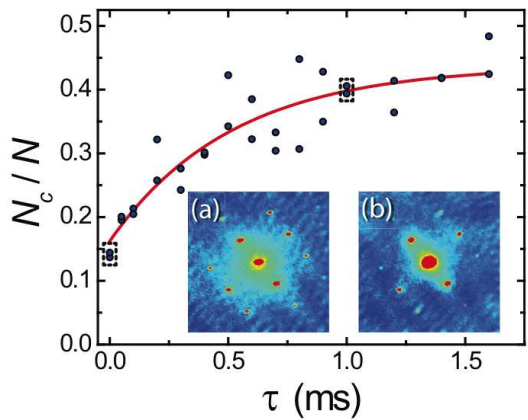}
\caption{\label{fig:data} Condensate fraction measured after bandmapping from $V_i=10$~$E_R$.  The insets show high optical density images where the background is resolved, but the Bragg peaks are saturated.   The images are shown in false color, with red (blue) indicating regions of high (low) column density.  The field of view is $813\times813$~$\mu$m.
(a)  $\tau=10$~ns;  (b)  $\tau=1$~ms. 
}
\end{figure}

Ten milliseconds after loading the lattice, we linearly ramp $V$ from $V_i$ to  $V_f=4$~$E_R$ in time $\tau$.  The lattice and trapping potentials are then removed in 10~ns and 0.2~ms, respectively, and the column density is imaged after the gas expands for 20~ms.   We extend the dynamic range of our measurement by imaging only a controlled fraction of atoms that are transferred to the $F=2$ hyperfine state.  The number of condensate atoms $N_c$ is measured using multimodal fits to ``low optical density'' images for which only a small number of atoms are transferred.  We supplement these with ``high optical density" images, for which all are transferred and imaged.   In these images, the broad non-condensate component is resolvable, but the condensate peaks are saturated (see Fig.~\ref{fig:data}).  The number of non-condensate atoms is determined by fitting the broad background with the condensed peaks masked and extrapolating the non-condensate component into the masked regions. The total number of particles $N$ varied from $\left(103\pm5\right)\times10^3$ to $\left(72\pm2\right)\times10^3$ for $V_i=10$~$E_R$ to 14~$E_R$, and the condensate fraction ranged from $0.3$ to $0.05$.

\begin{figure}
\begin{picture}(150, 140)
\put(-30, -5){\includegraphics[scale=0.5]{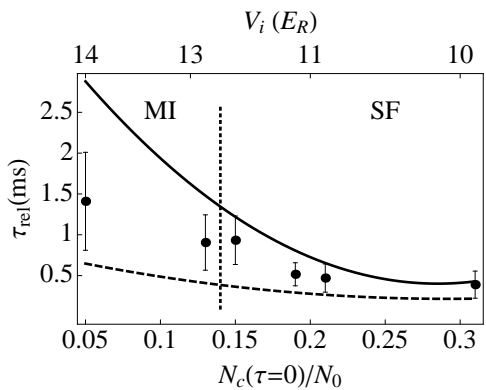}}
\end{picture}

\caption{\label{fig:results}Relaxation time ($\tau_{rel}$) for the condensate fraction for ramps from $V_{i}$ to $V_f=4E_{R}$ for variable ramp times $\tau$. 
The range of $V_{i}$ spans the superfluid (SF) and Mott insulator (MI) regimes (demarcated by the vertical dotted line). The experimental data is bounded by zero temperature Gutzwiller mean-field simulations using two different initial states (see text). The solid black line shows the relaxation time assuming initial state $2$, while the dashed line shows the relaxation time assuming initial state $1$. The error bars represent the uncertainty in the relaxation time from a fit to data such as that shown in Fig. \ref{fig:data}}
\end{figure}

\textit{Experimental Results.---} 
For each $V_i$ we measure the post-ramp condensate fraction as a function of $\tau$.  A typical data set is 
shown in Fig. \ref{fig:data} for $V_i=10$~$E_R$.  The data points follow an exponential, as illustrated by the red curve.  The fitted time constants $\tau_{\rm rel}$ are shown in Fig.~\ref{fig:results} as a function of the initial condensate fraction (bottom axis) along with the corresponding $V_i$ (top axis).  The relaxation time is relatively weakly dependent on $V_i$, only changing by a factor of two as the superfluid-Mott insulator transition is crossed, and throughout is consistent with the simple empirical rule $\tau_{\rm rel}\propto 1/U$.

The insets to Fig.~\ref{fig:data} shows the high optical density images for short (a) and moderate (b) ramps. Even for ramps as short as $1$ms, the quasi-momentum distribution changes dramatically, as atoms are transferred to low momentum. Most bandmapping experiments use $\sim 1$~ms for $\tau$, and therefore {\em do not} measure the initial condensate fraction.  Shorter ramps are not a solution to this problem, as they lead to significant non-adiabatic transfer of atoms to higher energy states \cite{zener}.

\textit{Theoretical Modeling.---} We use the $3$D Bose Hubbard Hamiltonian~\cite{fisher_boson_1989} to model our system:
\begin{equation}\label{eq:1}
{\cal{H}} = -\J\sum_{\langle ij \rangle}\left(a^{\dagger}_{i}a_{j} + h.c.\right) +\sum_{i}\left[ \frac{U}{2}n_{i}(n_{i}-1)
- \mu_i n_{i}\right]
\end{equation}
where $a_i$ and $a_i^\dagger$ are bosonic annihilation and creation operators at lattice site $i$, and $\mu_i = \mu - V_{\text{ex}}(i)$, where $\mu$ is the chemical potential and $V_{\text{ex}}(i)$ is the external potential at site $i$~\cite{jaksch_cold_1998}.  The first sum in Eq. 1 is over all nearest neighbor sites.  We calculate $U$ using the exact Wannier functions in the lowest band~\cite{zwerger_mott-hubbard_2003}, and extract the tunneling amplitudes $\J$ from Mathieu characteristics.   

We calculate dynamics using a time dependent Gutzwiller ansatz~\cite{fisher_boson_1989}, which approximates the wave-function by $\Psi = \bigotimes_{i}\sum_{m}c_{m}^{(i)}(\tb)|m\rangle_i$ where $|m\rangle_i$ is the $m$-particle Fock state on site $i$, and the coefficients $c_{m}^{(i)}(\tb)$ are space ($i$) and time ($\tb$) dependent. (Note the typographic distinction between the tunneling $\J$, time $\tb$, and ramp time $\tau$.) This approximation leads to a simplistic quasi-momentum distribution, dividing atoms into zero momentum ($k=0$) condensed and $k \neq 0$ non-condensed states.  The total number of condensed atoms $N_{c}$ is given by $N_{c} = \sum_{i}|\langle a_{i} \rangle|^2$, where $\langle a_{i} \rangle=\sum_m \sqrt{m+1}c_{m+1}^{(i)} \: c_m^{(i)}$.

As described in Ref.~\cite{natu}, Schr\"{o}dinger's equation $i\hbar\partial_{\tb}\psi = {\cal{H}}\psi$ for $\Psi$ yields a set of coupled differential equations for  the $c^{(i)}_{m}$:
\begin{eqnarray}\label{eq:2}
i\hbar\partial_{\tb}c^{(i)}_{m}(\tb) = -6\J(\tb)\left(\phi_{i}^{*}\sqrt{m+1}c^{(i)}_{m+1}  + \phi_{i}\sqrt{m}c^{(i)}_{m-1}\right) +\hspace{-6mm}\nonumber\\ \left[\frac{U(\tb)}{2}m(m-1) - \mu_{i} m\right]c^{(i)}_{m},
\end{eqnarray}
where the mean-fields are $\phi_{i} =  \sum_{\langle j \rangle}\langle a_{j} \rangle/6$, in which the sum is restricted to nearest neighbors  of site $i$. 

As in the experiment, the potential depth is ramped
 $V(\tb) = V_{i} +   (V_{f}-V_{i})(\tb/\tau)$, where $V_{i}$ and $V_{f}=4$~$E_R$ are the initial and final lattice depths, and $\tau$ is the ramp time.  The Hubbard parameters $\J$ and $U$ are time-dependent because of this ramp. In the simulations we include a spherically symmetric external harmonic trapping potential $V_{ex}$, matched to the (lattice-depth-dependent) experimental value $\bar{\nu}$.

We make direct comparison with the experimental data by studying $N \sim 75,000$ $^{87}$Rb atoms on a $55 \times 55 \times 55$ lattice with lattice spacing $406$~nm. For our initial state we use a local density approximation obtained by solving the homogenous, single-site problem. To account for the overestimation of the condensate fraction in mean-field theory, we use two different initial states to model the data.
{\em Initial state $1$} is the mean-field ground state obtained by using the physical lattice depth $V_{i}$ in the simulation. This state has a larger condensate fraction than the experimental initial state.  
{\em Initial state $2$} is obtained by finding the lattice depth 
at which the condensate fraction predicted by the theory
matches the measured condensate fraction at $V_{i}$.

Time evolution, from either of these initial states, is calculated using a split-step approach with sequential site updates. We ramp the lattice down from $V_{i}$ to $4E_{R}$ in a time $\tau$ ranging from $0$ to $1.5$~ms, and calculate the condensate fraction at the end of each ramp. We then fit the resulting data to an exponential curve to extract a characteristic relaxation time $\tau_{rel}$.

\textit{Comparison of Theory and Experiment --}  
In Fig.~\ref{fig:results} we compare two theoretical curves with the experimental data.

The bottom line is obtained from initial state $1$,  
and the top line is obtained from initial state 2.  
Initial state $1$ yields a \textit{higher} condensate fraction compared to the experimental system at any given initial lattice depth.  It relaxes to equilibrium faster, hence providing a \textit{lower} bound on the characteristic relaxation time $\tau_{rel}$.  Initial state $2$ treats the atoms as if they were in a deeper lattice than the physical system, and therefore leads to slower dynamics.  It therefore provides an \textit{upper} bound on the characteristic relaxation time. The dashed line indicates the lattice depth for which a shell of unit-filling Mott-insulator emerges.  Throughout, the data points lie between these two theoretical bounds.

For $V_i\lesssim13$~$E_R$ both theoretical protocols yield similar results.  Here the entire system is superfluid, and mean-field theory is accurate.  Throughout this regime $\tau_{\rm rel}\sim 0.5$ms.  As the Mott transition is approached, the relaxation time increases by a factor of $\sim$2, indicative of slower dynamics in the insulating state.  The simulation using initial state $2$ (top curve) captures this physics, showing a significant increase in relaxation time; initial state $2$ contains a Mott-insulator shell. The growth of the relaxation rate from initial state $1$ is more gradual, as the Mott-insulator transition occurs for larger values of $V_{i}$ as compared to the experiment.  While the slower timescales for dynamics in the Mott-insulator state are quite intuitive, the restoration of phase coherence following a rapid quench is not fully understood \cite{dziarmaga, Chen}.

Our simulations used zero temperature initial states --- finite temperature would modify the connection between condensate fraction and lattice depth, effectively raising the relaxation rate obtained from initial state $1$.  

\begin{figure}
\begin{picture}(150, 80)
\put(-50, -10){\includegraphics[scale=0.335]{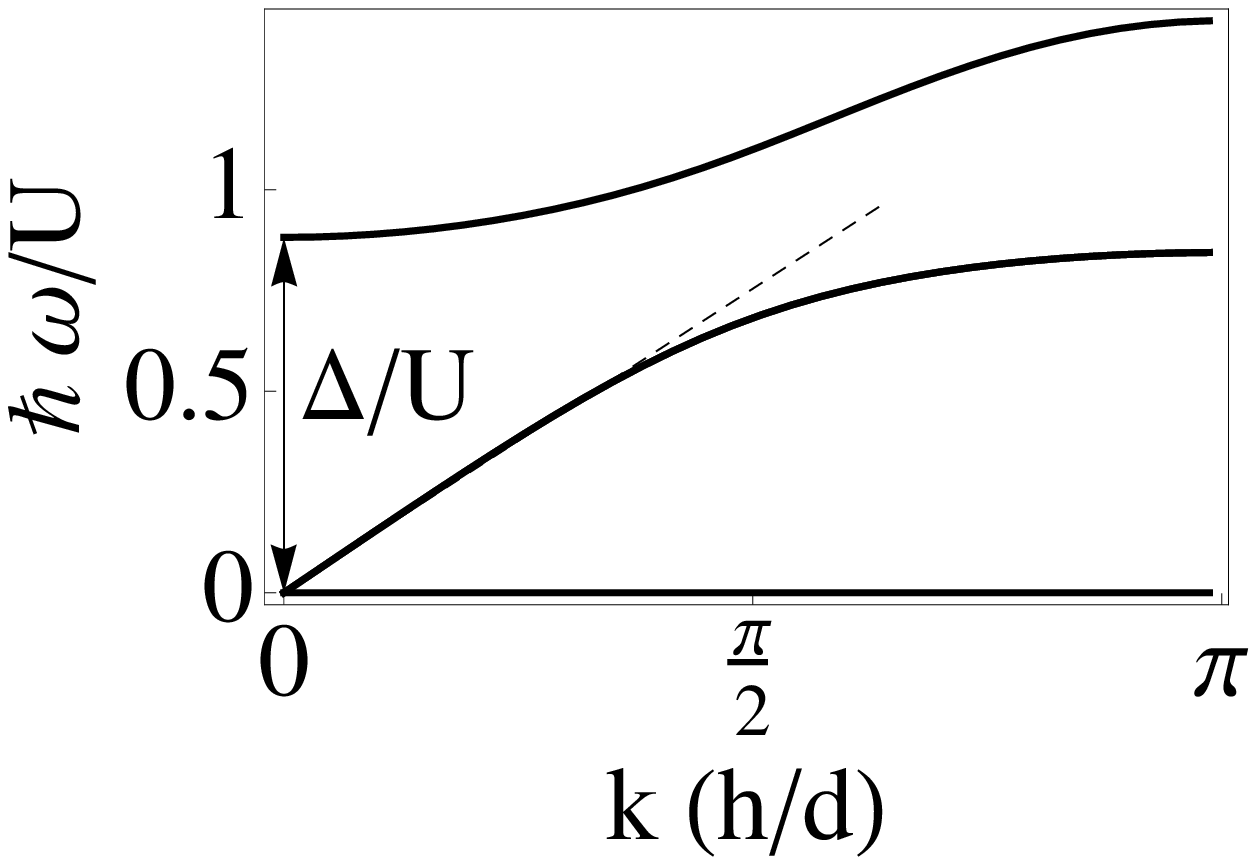}}
\put(75, -10){\includegraphics[scale=0.335]{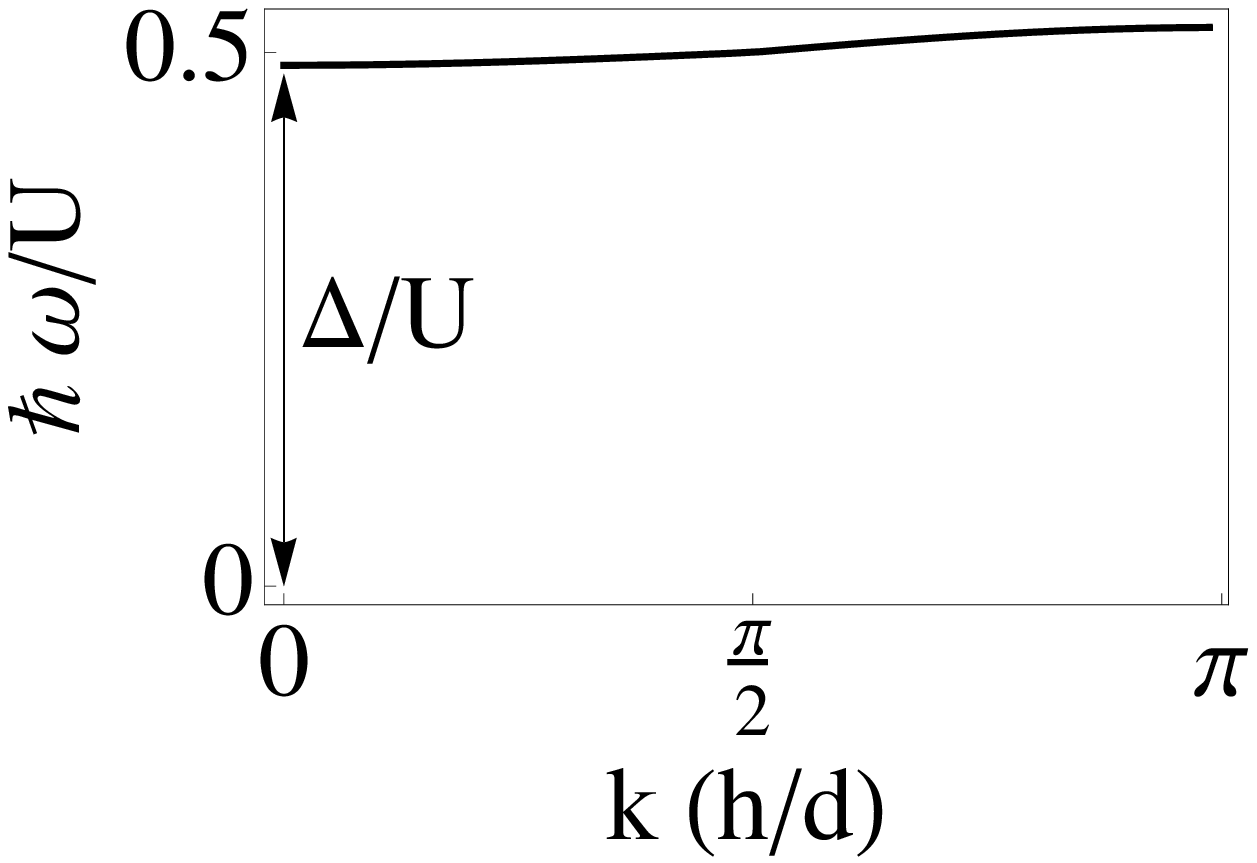}}
\end{picture}

\caption{\label{fig:-2}Typical Gutzwiller excitation spectra for deep lattices. (Left): \textit{Quantum depleted superfluid near the Mott-insulator transition} ($V = 13E_R$). Two modes are present: a gapless phonon mode with a linear Bogoliubov (dashed) dispersion at low $k$, and a gapped particle-hole mode with quadratic dispersion. The gap $\Delta \sim U$, sets the diabaticity timescale for bandmapping. (Right): \textit{Mott insulator at $\mu = U/2$ and $V = 28E_R$}.  The lowest energy excitation has a gap at $k=0$ of order $U/2$, that sets the diabaticity time.}
\end{figure}

\textit{Understanding the timescales.---} Bandmapping timescales are too short for any significant transport in the lattice to occur \cite{natu}, and thus the observed physics is purely local. One gains insight by considering the homogenous case 
 $V_{ex} = 0$. Linearizing Eq.~\ref{eq:2} about the homogenous Gutzwiller stationary state \cite{krutitsky}, we find the excitation spectrum $\omega(k)$ for the lattice gas in the shallow and deep lattice limits (Fig.~\ref{fig:-2}). 

In the shallow lattice, the only relevant excitations are linearly dispersing phonons. Excitations that change the overall phase of the wave-function cost no energy (c.f Fig.\ref{fig:-2} (left)). On short timescales, the overall density remains constant and phonons are not excited.  For deeper lattices, a particle-hole branch with a quadratic dispersion, and a gap $\Delta$ is present.   

The superfluid state in a deep lattice is ``quantum depleted": the fraction of condensed ($k = 0$) atoms approaches $0$ near the insulating transition. Upon lowering the lattice depth quasi-momentum states coherently evolve from high momentum to low momentum. Full phase coherence is restored when these excited atoms subsequently decay into particle-hole pairs of comparable energy but low  momentum, while changing the phase of the local wave-function. The timescale for this process is set by $1/\Delta$, the timescale to excite a particle-hole excitation. 

We produce an analytic expression for $\Delta$ by truncating the Fock basis to at most $2$ particles per site and diagonalizing the resulting $3 \times 3$ Hamiltonian. The eigenvalues become particularly simple when $\mu \approx U/2$, and we find that $\Delta = \frac{1}{4}(U + \sqrt{48 z^2\langle a \rangle^2t^2 + U^2})$, where $z$ is the co-ordination number ($z=6$ for a cubic lattice). At $\J \rightarrow 0$ and unity filling, the gap $\Delta \rightarrow U/2 \sim 
1.5$ms in deep lattices (Fig.~\ref{fig:-2}). Therefore, even in the superfluid phase, interactions foil bandmapping because it is not possible to ramp off the lattice quickly compared to $h/\Delta$ and slowly compared with $h/E_{bg}$.

\textit{Summary.---} We find that previous assumptions that dynamics are frozen during bandmapping are incorrect for interacting systems. Rather, we observe that considerable momentum redistribution occurs for typical bandmapping times, driven by the presence of particle-hole excitations in the quantum depleted superfluid. These excitations render bandmapping unreliable for measuring quasi-momentum distributions in interacting systems. This problem could be circumvented by turning off interactions prior to bandmapping using a Feshbach resonance. 

We remark that much of the physics of these ramps is captured by time-dependent Gutzwiller mean-field theory.  There is, however, room for quantitative improvement: for ramps from deeper lattices, different assumptions lead to a five-fold variation in the relaxation time $\tau_{rel}$.  To accurately model the experimental findings, as well as answer subtler questions, such as how correlations propagate across the gas, one must resort to more sophisticated techniques. \cite{blochcor}. 

Extracting information about the underlying many body physics from quench experiments is a vibrant area of research \cite{blochcor, polkovnikov}, whose importance transcends ultra-cold atoms. Here we have shown that the timescales governing the evolution of the condensate fraction during a lattice ramp is an indirect measure of particle-hole excitations. Probes such as bandmapping are therefore indispensible to developing our understanding of strongly correlated systems. 

\textit{Acknowledgements.---} We thank Randy Hulet and Joseph Thywissen for enlightening discussions. This work was supported by a grant from the Army Research Office with funding from the DARPA OLE program.

\end{document}